\documentclass[aip,reprint,showpacs,floatfix,amsmath,amssymb]{revtex4-1}

\usepackage{graphicx, color}
\usepackage{amsmath,amssymb,wasysym}
\usepackage{float}
\usepackage{verbatim}
\usepackage{subfigure}
\usepackage{sidecap}
\usepackage{multirow}
\usepackage{marginnote}
\definecolor{goldenrod}{rgb}{0.85, 0.65, 0.13}

\newcommand{\figOne}{
 \begin{figure}[!ht] \hspace*{-0cm} 
\centering
    \includegraphics[scale=.4]{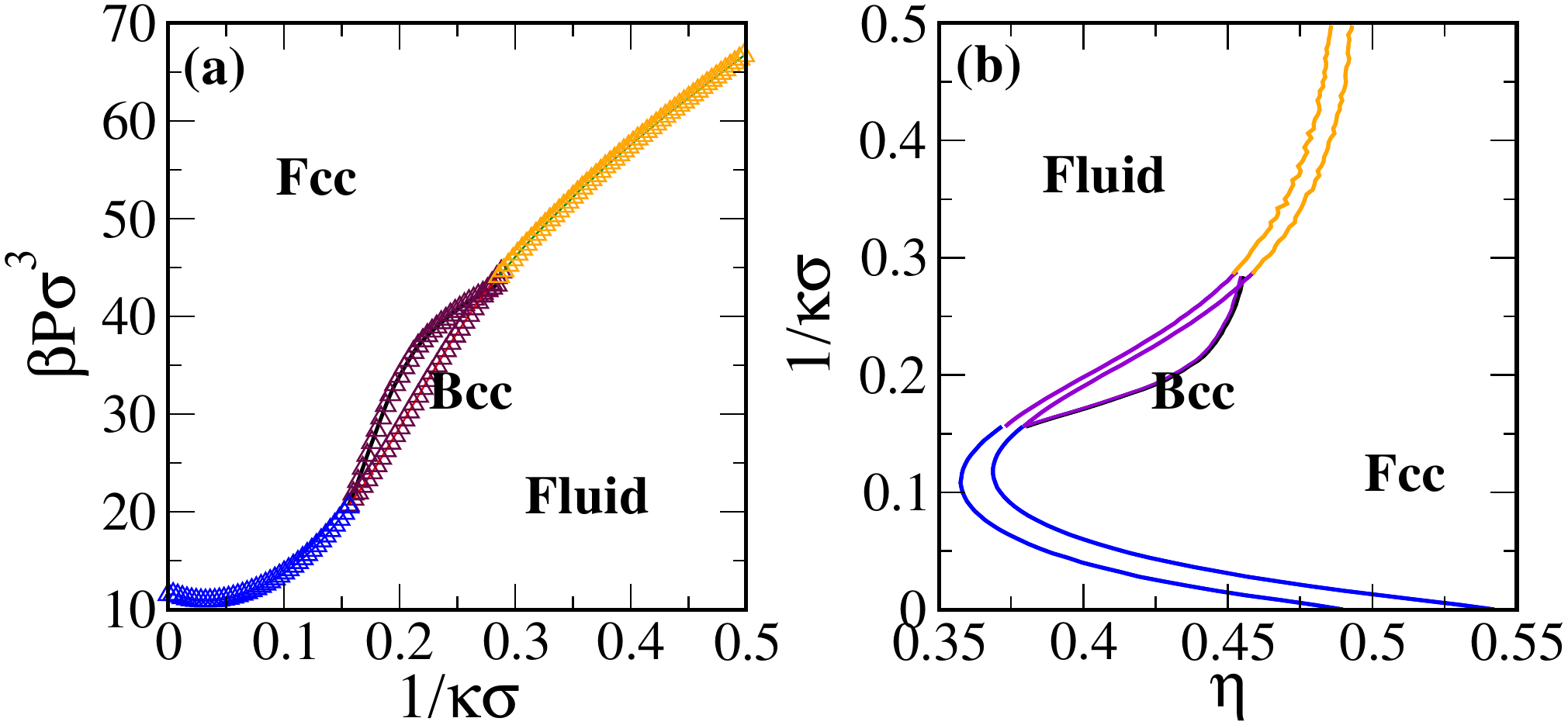}
 \caption{Phase diagram of a system in which the particles interact 
 via a hard-core repulsive Yukawa pair potential with $\beta\epsilon =8$ in \textbf{(a)} the $(1/\kappa\sigma-\beta P\sigma^3)$ representation and \textbf{(b)} the $(1/\kappa\sigma-\eta)$ plane. The phase diagram displays a stable fluid, bcc, and fcc phase. }\label{fig:101}
\end{figure}
}

\newcommand{\figTwo}{
\begin{figure}[!htbp]
\centering
  \includegraphics[scale=0.35]{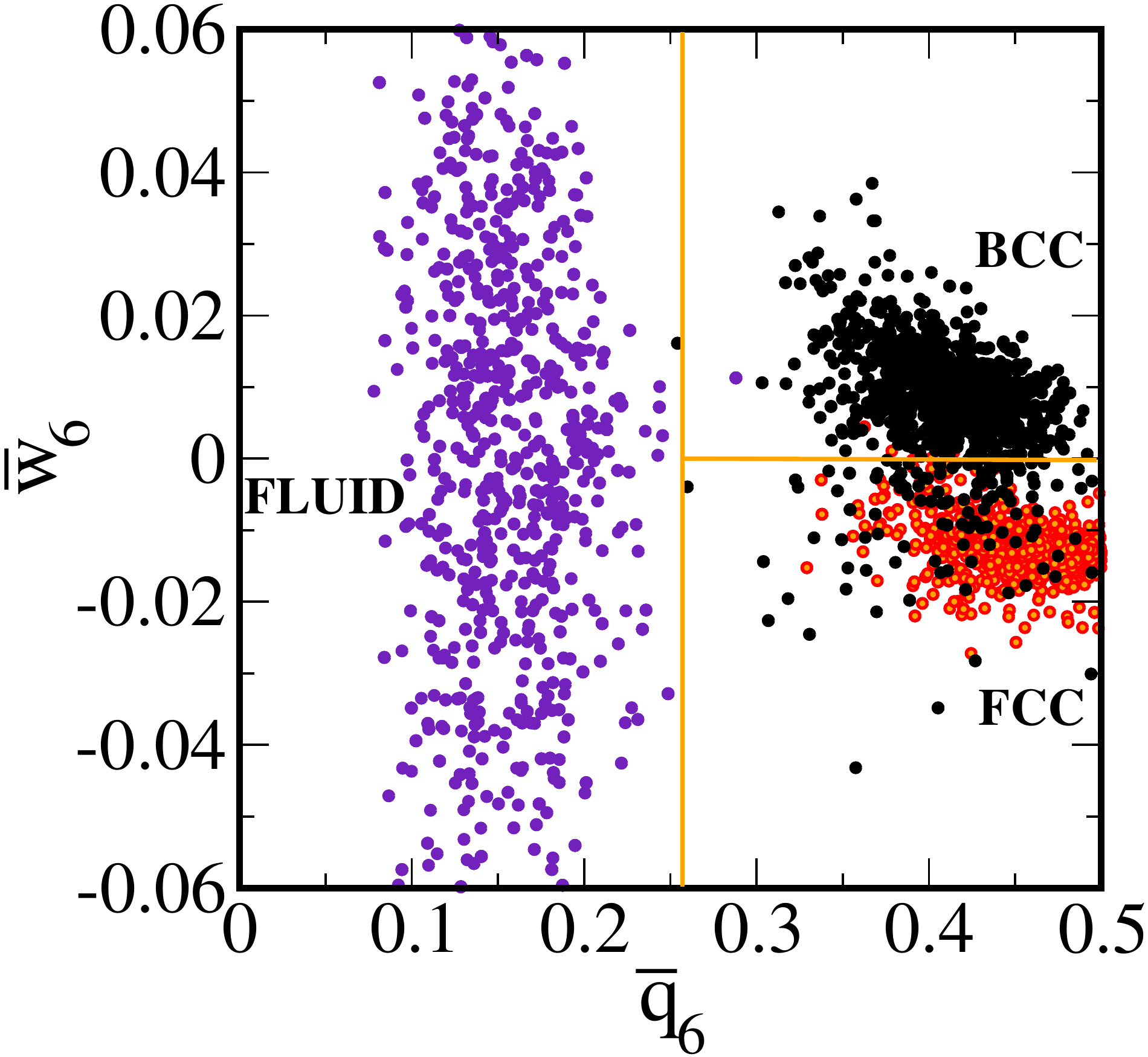}
\caption{Scatter plot of the averaged bond order parameters  $\bar{q}_6$ {\it versus} $\bar{w}_6$ for the fluid, bcc, and fcc  phase of a system of Yukawa particles with contact value $\beta \epsilon=8$ at the high-density triple point. Each point corresponds to a single particle. In total  2000 points  were chosen randomly from each structure. }\label{fig:q411}
\end{figure} 
}

\newcommand{\figThree}{
 \begin{figure*}[]
\centering
  \includegraphics[scale=0.6]{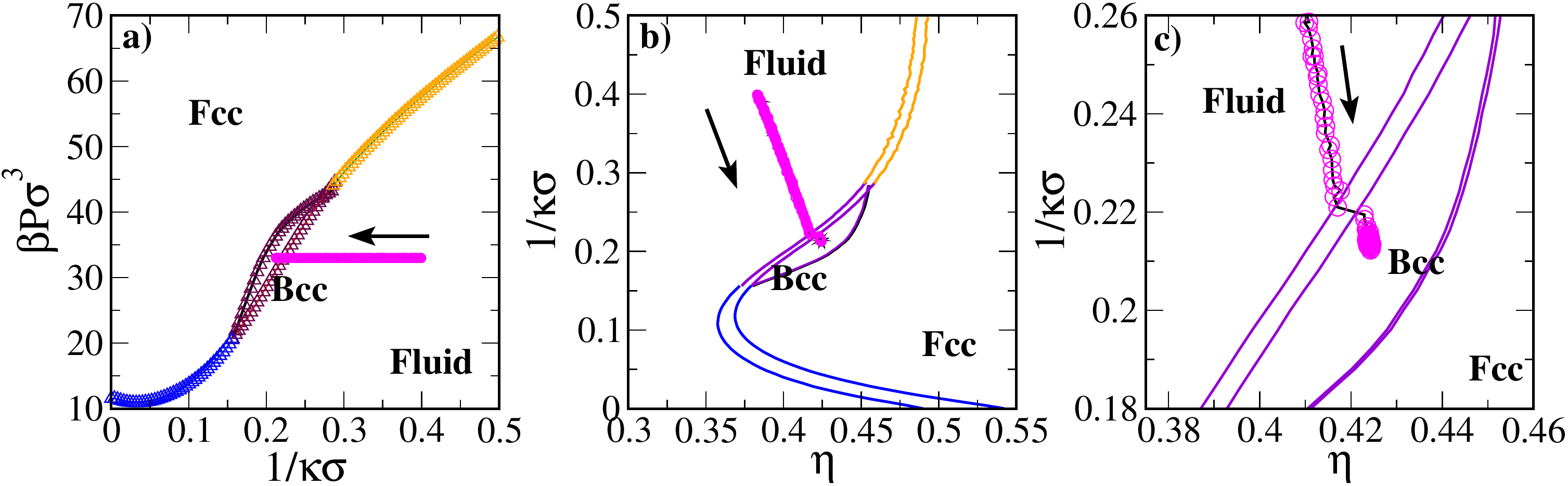}

\caption{Evolution of the parameters in \textbf{(a)} the ($1/\kappa\sigma-\beta P\sigma^3$) plane and  \textbf{(b)} the ($1/\kappa\sigma-\eta$) plane when the system is initialized  in the fluid phase at pressure $\beta P\sigma^3=33$, inverse Debye screening length $1/\kappa\sigma=0.4$ and contact value $\beta\epsilon=8$. $\beta P\sigma^3$ and $\beta\epsilon$ are kept fixed, and only $1/\kappa\sigma$ is tuned.  \textbf{(c)} shows an enlarged view of the data near the fluid-bcc coexistence region.}\label{denFluid132}
\end{figure*}
}

\newcommand{\figFive}{
\begin{figure*}[!ht]
\centering
  \includegraphics[scale=0.6]{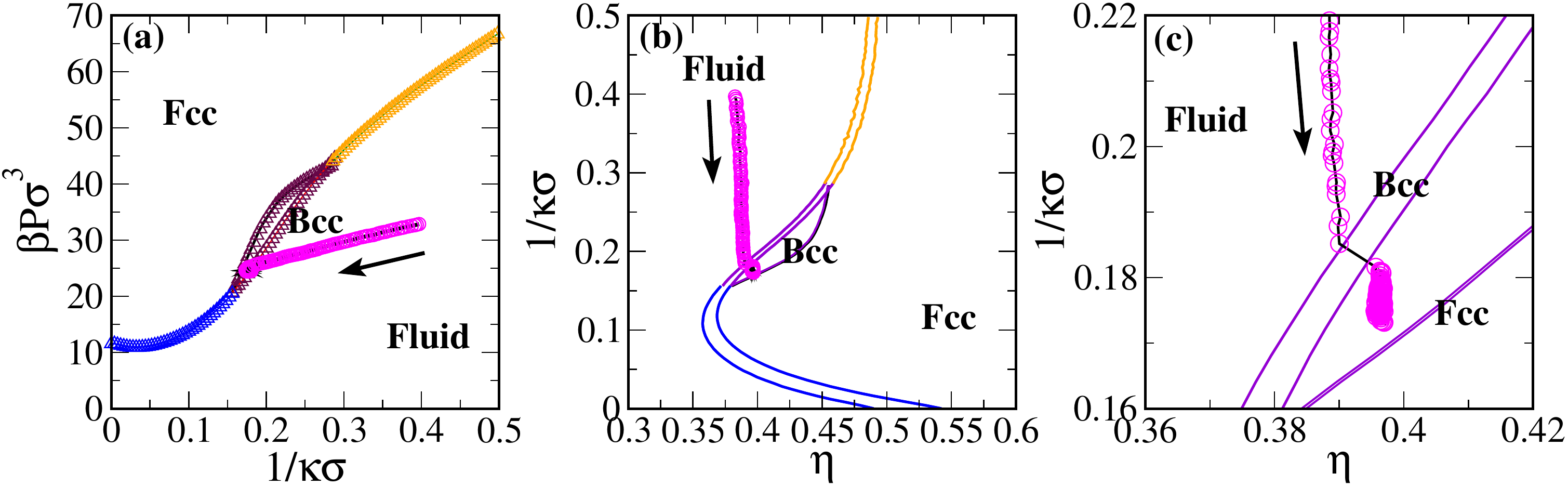}
  
\caption{Evolution of the parameters in \textbf{(a)} the ($1/\kappa\sigma-\beta P\sigma^3$) plane and \textbf{(b)} the ($1/\kappa\sigma-\eta$) plane when the system is initialized  in the fluid phase at pressure $\beta P\sigma^3=33$, inverse Debye screening length $1/\kappa\sigma=0.4$ and contact value $\beta\epsilon=8$.  $\beta\epsilon$ is kept fixed, and two parameters $1/\kappa\sigma$ and $\beta P\sigma^3$ are tuned.  \textbf{(c)} shows an enlarged view of the data near the fluid-bcc coexistence region.}\label{fig:a09}
\end{figure*}
}

\newcommand{\figFour}{
\begin{figure*}[!htbp]
\centering
  \includegraphics[scale=0.6]{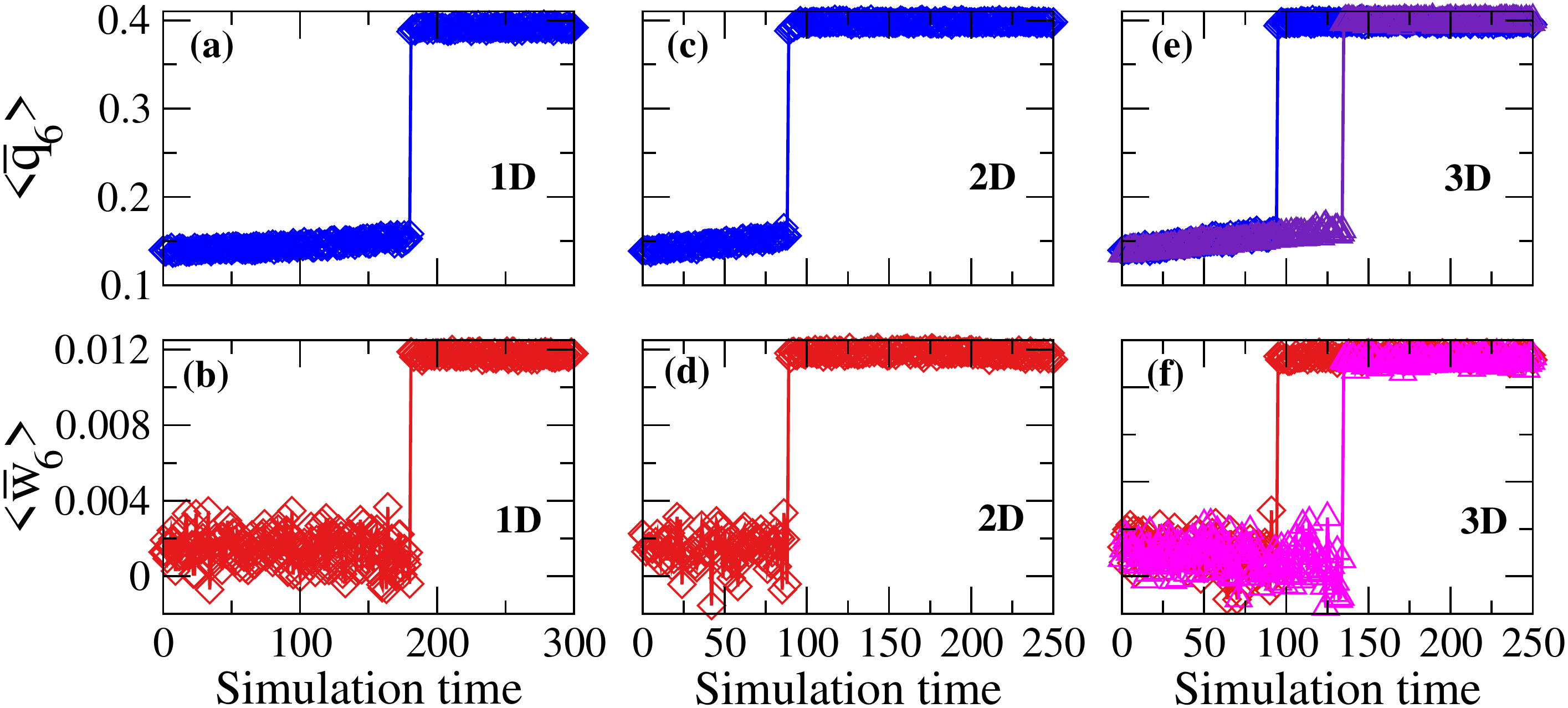}

\caption{Evolution of $\langle\bar{q}_6\rangle$ and $\langle\bar{w}_6\rangle$ as a function of simulation time during the optimization using the SP-ID algorithm for  \textbf{(a-b)} the 1$D$, \textbf{(c-d)} the 2$D$, and \textbf{(e-f)} for the 3$D$ case, respectively.  The system is initialized  in the fluid phase at pressure $\beta P\sigma^3=33$, inverse screening length $1/\kappa\sigma=0.4$ and contact value  $\beta\epsilon=8$. For the $3D$ case, blue and red colored symbols represent the $\langle\bar{q}_6\rangle$ and $\langle\bar{w}_6\rangle$ values when the system is initialized at (i) $1/\kappa\sigma=0.4$, $ \beta P \sigma^3=33$, $\beta\epsilon=8$, while indigo and magenta colored symbols represent bond order parameter values for the case (ii) $1/\kappa\sigma=0.4$, $ \beta P \sigma^3=25$, $\beta\epsilon=6$. The simulation time indicates the number of simulations performed at distinct sets of interaction parameters.  }\label{fig:bop}
\end{figure*}
}

\newcommand{\figSix}{

 \begin{figure*}[!ht]\hspace*{-0cm}
\centering
  \includegraphics[scale=.6]{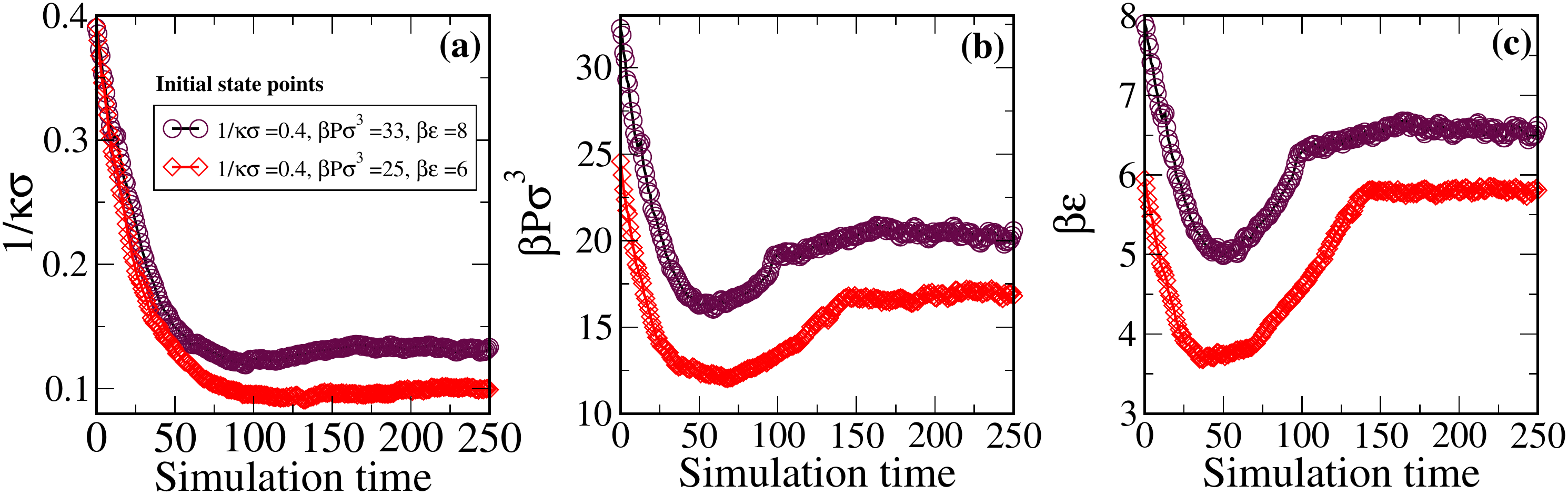}
  
\caption{ Evolution of \textbf{[(a)]} $1/\kappa\sigma$, \textbf{[(b)]} $\beta P \sigma^3$, and \textbf{[(c)]} $\beta\epsilon$ as a function of simulation time when the system is initialized  in the fluid phase at (i)  $\beta P\sigma^3=33$, $1/\kappa\sigma=0.4$  $\beta\epsilon=8$ (indigo colored circles), and (ii)  $\beta P\sigma^3=25$, $1/\kappa\sigma=0.4$  $\beta\epsilon=6$ (red colored diamonds). Three parameters are tuned to target the bcc structure, namely $\beta P \sigma^3$, $1/\kappa\sigma$ and $\beta\epsilon$. 
}\label{fig:13}
\end{figure*}
}

\newcommand{\tableOne}{
\begin{table}[!ht]
\centering 
	\def\arraystretch{1.25}
 	\setlength\arrayrulewidth{1.25pt}
\begin{tabular}{|c|c|c|c|}
 
\hline

                                                                              & $1/\kappa\sigma$ & $\beta P\sigma^3$   & $\beta\epsilon$ \\
\hline                                            

\begin{tabular}[c]{@{}l@{}}Initial state point\\   (fluid phase)\end{tabular} & 0.4              & 33.0  & 8.0             \\
\hline
\begin{tabular}[c]{@{}l@{}}Final state point\\   (bcc phase)\end{tabular}      & 0.13             & 20.0 & 6.5    	   \\ 
\hline
\end{tabular}
\caption{The initial state point values corresponding to the fluid phase in the phase diagram as given to the optimization algorithm and the final state point values obtained by the optimization algorithm  corresponding to the bcc phase.
 			 }\label{table7}
\end{table}
}

\newcommand{\figSeven}{
\begin{figure}[!htbp]
\centering
  \includegraphics[scale=0.76]{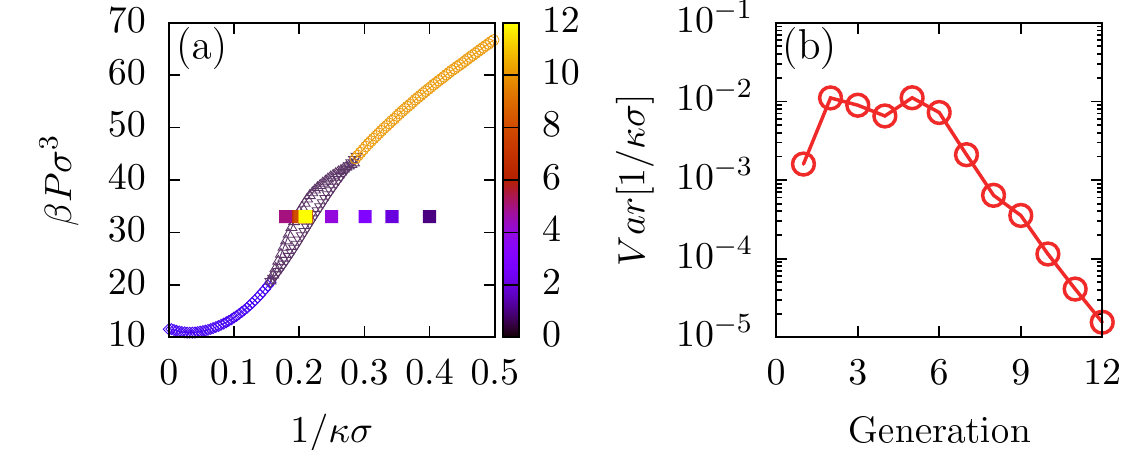}

\caption{\textbf{(a)} Evolution of the the mean value of the  Gaussian distribution for $1/\kappa\sigma$ in  the ($1/\kappa\sigma-\beta P\sigma^3$) plane  when the system is initialized  in the fluid phase at pressure $\beta P\sigma^3=33$, inverse Debye screening length $1/\kappa\sigma=0.4$ and contact value $\beta\epsilon=8$.  The parameters $\beta\epsilon$ and $\beta P\sigma^3=33$ are kept fixed, and  $1/\kappa\sigma$ is tuned using the CMA-ES method.  \textbf{(b)} The  variance of the Gaussian distribution for $1/\kappa\sigma$ at each generation.}\label{fig:cma1}
\end{figure}
}
\newcommand{\figEight}{
\begin{figure}[!htbp]
\centering
  \includegraphics[scale=1.1]{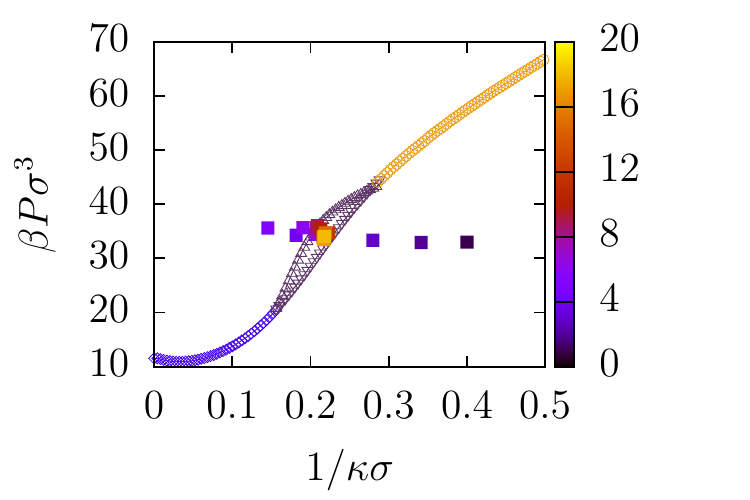}

\caption{Evolution of the the mean value of the  multivariate Gaussian distribution for $1/\kappa\sigma$ and $\beta P\sigma^3$ in  the ($1/\kappa\sigma-\beta P\sigma^3$) plane  when the system is initialized  in the fluid phase at pressure $\beta P\sigma^3=33$, inverse Debye screening length $1/\kappa\sigma=0.4$ and contact value $\beta\epsilon=8$.  The parameter $\beta\epsilon$ is kept fixed, and  $1/\kappa\sigma$ and $\beta P\sigma^3$ are tuned using the CMA-ES method. }\label{fig:cma2}
\end{figure}
}
\newcommand{\figNine}{
\begin{figure*}[!htbp]
\centering
  \includegraphics[scale=0.6]{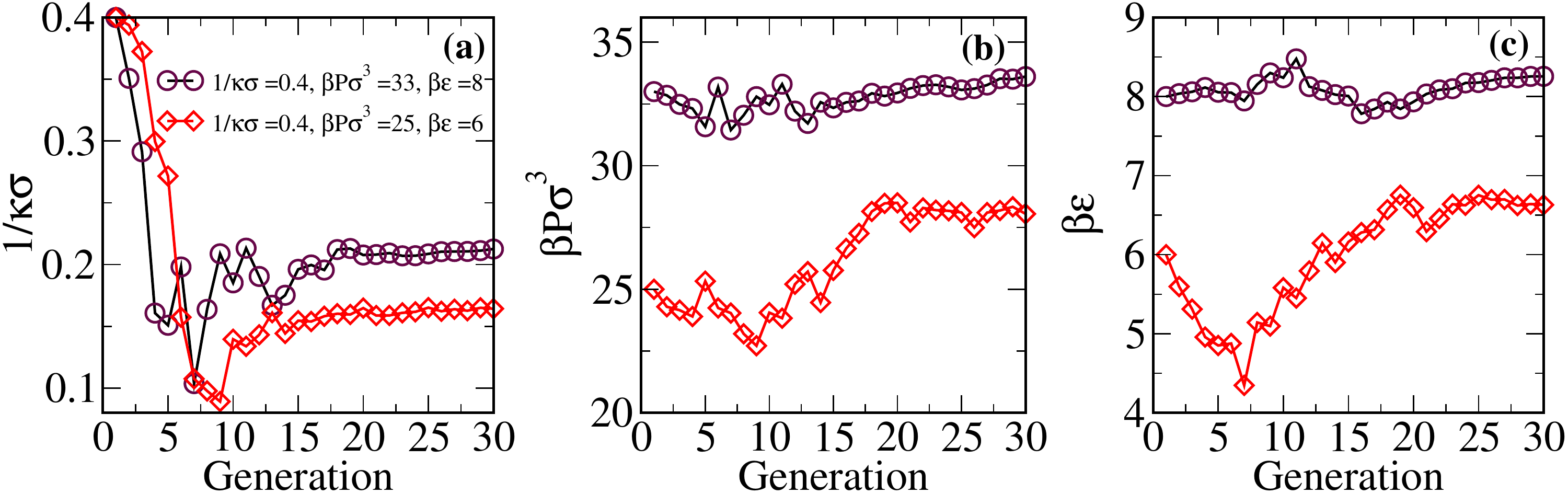}

\caption{Evolution of the the mean value of the  multivariate Gaussian distribution for inverse Debye screening length $1/\kappa\sigma$, pressue $\beta P\sigma^3$ and contact value $\beta\epsilon$  when the system is initialized  in the fluid phase at (i)  $\beta P\sigma^3=33$, $1/\kappa\sigma=0.4$  $\beta\epsilon=8$ (indigo colored circles), and (ii)  $\beta P\sigma^3=25$, $1/\kappa\sigma=0.4$  $\beta\epsilon=6$ (red colored diamonds). Three parameters are tuned to target the bcc structure, namely $\beta P \sigma^3$, $1/\kappa\sigma$ and $\beta\epsilon$ using the CMA-ES method.
}\label{fig:cma3}
\end{figure*}
}
\newcommand{\figTen}{
\begin{figure}[!htbp]
\centering
  \includegraphics[scale=0.5]{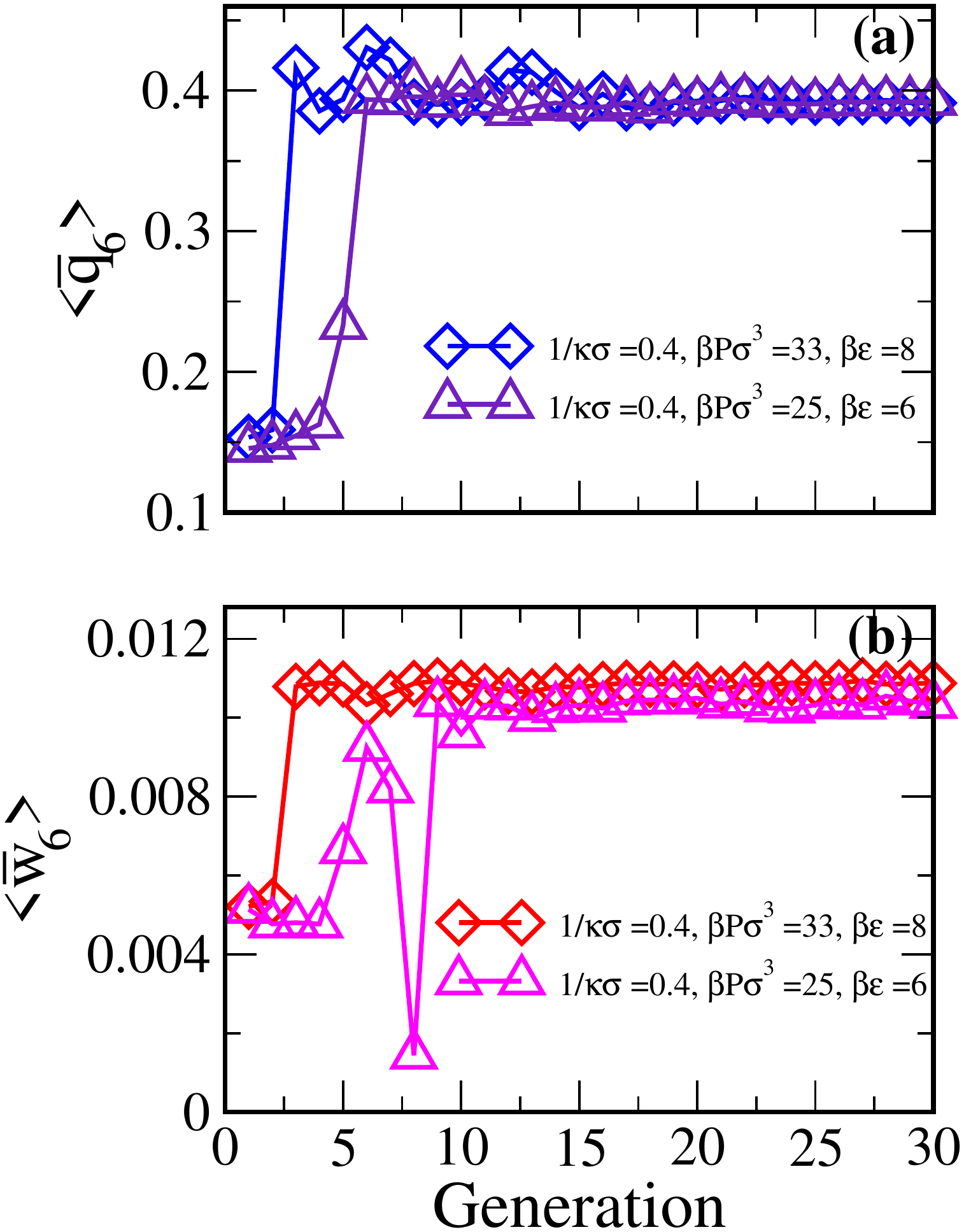}

\caption{Evolution of \textbf{(a)} $\langle\bar{q}_6\rangle$ and \textbf{(b)} $\langle\bar{w}_6\rangle$ of the sample with the highest value of the fitness at each generation during the 3$D$ optimization with the CMA-ES method for both the investigated cases. We observe a jump of $\bar{q}_6$ and $\bar{w}_6$ when, for the first time, at least one of the generated samples is in the bcc region, even if the mean value of the multivariate Gaussian distribution at the same generation does not lie in the same region.
}\label{fig:cmabop}
\end{figure}
}

\definecolor{cream}{RGB}{222,217,201}

\begin{document}

\title{Inverse design of charged colloidal particle interactions for self assembly into specified crystal structures}
\author{Rajneesh Kumar}
\email{rajneesh@jncasr.ac.in}
\affiliation{Theoretical Sciences Unit, Jawaharlal Nehru Centre for Advanced Scientific Research, Bengaluru 560064, India.}
\author{Gabriele M. Coli}
\email{g.m.coli@uu.nl}
\affiliation{Soft Condensed Matter, Debye Institute for Nanomaterials Science, Utrecht University, Princetonplein 5, 3584 CC Utrecht, The
Netherlands}

\author{Marjolein Dijkstra}
\email{m.dijkstra@uu.nl}
\affiliation{Soft Condensed Matter, Debye Institute for Nanomaterials Science, Utrecht University, Princetonplein 5, 3584 CC Utrecht, The
Netherlands}

\author{{Srikanth Sastry}}
\email{sastry@jncasr.ac.in}
\affiliation{Theoretical Sciences Unit, Jawaharlal Nehru Centre for Advanced Scientific Research, Bengaluru 560064, India.}

\begin{abstract}
We study the inverse problem of tuning interaction parameters between charged colloidal particles interacting with a hard-core repulsive Yukawa potential, so that they assemble into specified crystal structures.  Here, we target the body-centered-cubic (bcc) structure which is only stable in a small region in the phase diagram of charged colloids and is, therefore, challenging to find. In order to achieve this goal, we  use the statistical fluctuations in the bond orientational order parameters to tune the interaction parameters for the bcc structure, while initializing the system in the fluid phase, using the Statistical Physics-inspired Inverse Design (SP-ID) algorithm~\cite{Miskin}. We also find that this optimization algorithm correctly senses the fluid-solid phase boundaries for charged colloids. Finally, we repeat the procedure employing the Covariance Matrix Adaptation - Evolution Strategy (CMA-ES), a cutting edge optimization technique, and compare the relative efficacy of the two methods. 
	

\end{abstract}


\maketitle

\section{Introduction} \label{sec:I}
In the past few years, several inverse methods have emerged that design optimal interactions in such a way that the system spontaneously assembles into a targeted structure~\cite{Rechtsman0,Rechtsman1,Rechtsman2,Rechtsman3}.
These methods have received considerable attention in materials  science~\cite{Torquato,Jain,Jaeger}, and have been successively used to find crystal
 structures for photonic band-gap applications~\cite{Ho}, to predict crystals~\cite{Oganov} and protein structures~\cite{Dahiyat, Kuhlman}, materials with optimal mechanical and transport properties~\cite{Hyun}, and for optimizing the interactions for self assembly~\cite{Hannon0,Chang,Rechtsman4,Hannon1,Hormoz}. Though many of the methods developed are either based on  black-box techniques, in which the algorithm tunes the interaction parameters without taking the statistical nature of the system into account, or are designed \emph{ad hoc} for a particular class of systems, systematic approaches based on a statistical mechanical formulation, which are general, and allow for application tailored to specific systems of interest, have also been investigated.

In the present work, we investigate the efficacy of one such method, the Statistical Physics-inspired Inverse Design (SP-ID) method  developed by Miskin {\it{et al}} ~\cite{Miskin}. This method considers statistical fluctuations present in the microscopic configurations of the system for tuning the interactions between the particles. In order to design these interactions, we have used a  quality function based on bond order parameters~\cite{Stein,Lechner} to rank the generated configurations. The same task can be faced employing numerous optimization techniques like (Adaptive) Simulated Annealing ~\cite{Kirkpatrick,Ingber1,Ingber2}, Particle Swarm Optimization~\cite{DeOca}, and several genetic algorithms~\cite{Mitchell}. To evaluate the effectiveness of the SP-ID  algorithm, we compare it with the Covariance Matrix Adaption - Evolutionary Strategy (CMA-ES)~\cite{Hansen2001,Hansen2003,Hansen2006}, which we regard as a state-of-the-art optimization technique for evolutionary computation.

We have chosen a system  of colloidal particles as the model for which we wish to design the interactions in order to target a specific crystal structure. The interparticle potential of colloids offers a wide variety of functional forms. It can contain a hard-core term, a dipole-dipole term, a charge dispersion term, a screened-Coulomb (Yukawa) term and a short-ranged attractive depletion term. More specific designed colloidal particles such as patchy colloids and DNA-functionalized colloids offer even greater diversity of interactions. For all of these, the interaction  parameters can be  tuned. In the case of Yukawa interactions, the Debye screening length can be adjusted by changing the salt concentration, and the contact value can be tuned by altering the surface charge of the particles.  Previous studies have employed what may be termed a {\it forward method} in which, starting from the microscopic interaction parameters and specified thermodynamic parameters such as temperature and density, one computes the equilibrium properties and finds the stable phase of the system. In the present work, our purpose is to employ an {\it inverse method}, where  the target structure and equilibrium properties are the input from which we wish to design the interparticle interactions for which the particles spontaneously self assemble into the target structure.

In this study, we considered a charged colloidal system in which particles interact with a hard-core repulsive Yukawa potential.  The complete phase diagram of hard-core Yukawa particles is known from earlier studies~\cite{Meijer1,Meijer, Dijkstra}. Because of the purely repulsive nature of the potential, this system displays only a fluid phase which can freeze into a face-centered-cubic (fcc) or a body-centered-cubic (bcc) crystal phases.
The phase diagram of hard-core Yukawa particles shows two triple points where fcc, bcc, and fluid phases coexist. The bcc phase is only stable in a very small region in the phase diagram and, for this reason, constitutes a reasonable test case for the reverse-engineering process. In this work, we search for optimal interparticle interactions which favor  the crystallization of the bcc structure. Here we show three different cases, in which we respectively tune one, two and three parameters. In all  cases, we show that both SP-ID and CMA-ES adjust the interparticle interactions that lead to the targeted bcc structure formation.

The paper is organized as follows: In Sec.~\ref{sec:model}, we define the model system  studied in this paper and the corresponding phase diagram and  bond order parameters to identify the different phases. In Sec.~\ref{sec:ID}, we define the inverse design methods and the form of the quality function used in this study. Results for  tuning the interactions to target the bcc structure for all three cases are discussed in Sec.~\ref{sec:3}.
Finally, we summarize our results in Sec.~\ref{sec:4}.

\section{Model and Simulation Methods} \label{sec:model}
\textbf{Interaction Potential: }We consider a hard-core repulsive Yukawa system which represents a standard model for charged colloids. The form of the potential is given by
 \begin{equation}\label{eq:pot}
				\beta U(r) = 
				\begin{cases}
					\beta\epsilon\frac{\text{exp} \left[-\kappa\sigma \left(r/\sigma -1\right)\right]}{r/\sigma} & \text{for} \ r > \sigma \cr
					\infty  & \text{for} \ r \leq \sigma
				\end{cases}
			\end{equation}
	where $\beta\epsilon$ is the contact value of the pair potential expressed in units of
$k_BT=1/\beta$, $\kappa$ is the inverse of the Debye screening length, and $\sigma$ is the
hard-core diameter. In Fig.~\ref{fig:101}, we show the phase diagram for such a system with $\beta\epsilon =8$ in the $1/\kappa\sigma-\eta$ representation and the $1/\kappa\sigma-\beta P\sigma^3$ plane~\cite{Dijkstra}. The phase diagram exhibits a stable fluid, bcc, and fcc  region, as well as two triple points at which the three phases coexist. Note that we always use scaled variables, a reduced temperature $k_BT/\epsilon$, pressure $\beta P\sigma^3$ and inverse screening length $1/\kappa\sigma$, each of which can be treated as tuning parameter for  obtaining the desired behavior.\\

\textbf{Bond order parameters (BOP): }
Every optimization algorithm, including SP-ID and CMA-ES, works on the basis of minimizing a user-defined fitness function. Here, we have used the averaged bond order parameters $\bar{q}_l$ and $\bar{w}_l$ ($l=6$). The first is computed as follows:~\cite{Stein,Lechner},
 \begin{equation}
 \bar{q}_{l}^{(i)}=[\frac{4\pi}{2l+1}\sum\limits_{m=-l}^{l}|\bar{q}_{lm}^{(i)}|^2]^{1/2}
\end{equation}
where,
\begin{equation*}
 \bar{q}_{lm}^{(i)}=\frac{1}{\tilde{N}_b(i)}\sum\limits_{j=0}^{\tilde{N}_b(i)}q_{lm}^{(j)}
\end{equation*}
\begin{equation*}
 q_{lm}^{(i)}=\frac{1}{N_b(i)}\sum\limits_{j=1}^{N_b(i)}Y_{lm}(\theta(\mathbf{r}_{ij}),\phi(\mathbf{r}_{ij})).
\end{equation*}
Here,  $N_b(i)$ is the number of neighbors of particle $i$, $\tilde{N}_b(i)$ is the number of neighbors  including particle $i$ itself,   $Y_{lm}(\theta(\mathbf{r}_{ij}),\phi(\mathbf{r}_{ij}))$ denotes the spherical harmonics with $\mathbf{r}_{ij}$ the distance vector from particle $i$ to particle $j$.
The second bond order parameter we used is defined as 
$$\bar{w}_l^{(i)}= \frac{\sum\limits_{m_1+m_2+m_3}\begin{pmatrix}
l & l & l\\ 
m_1 & m_2 & m_3 
\end{pmatrix}  \bar{q}_{lm_1}^{(i)} \bar{q}_{lm_2}^{(i)}\bar{q}_{lm_3}^{(i)} }{(\sum\limits_{m=-l}^{l}|\bar{q}_{lm}^{(i)}|)^{3/2}}$$
\figOne

%

In order to calculate the radius of the first coordination shell for each particle, we employ the solid angle based nearest-neighbor (SANN) algorithm~\cite{Sann}, where a nearest neighbor of a particle is identified by attributing a solid angle to each possible neighbor such that the sum of  solid angles equals at least $4\pi$.

 In Fig.~\ref{fig:q411}, we show the scatter plots of  $\bar{q}_6$ {\it versus} $\bar{w}_6$ 
 for the fluid, bcc, and fcc phases of a system of Yukawa particles with $\beta \epsilon=8$ at the high-density triple point conditions. We find distinct clouds of points corresponding to the fluid, bcc, and fcc phases, and hence, $\bar{q}_6$ and $\bar{w}_6$ can be used to distinguish the three phases. 
 
\section{Inverse design Methods}\label{sec:ID}
\subsection{Statistical Physics-inspired Inverse Design method}
In the SP-ID method, microscopic parameters such as the interparticle pair potential parameters  are tuned by exploiting statistical fluctuations in such a way that the system will evolve to those states which correspond to the targeted macroscopic response of that system, unlike other models which are entirely based on black-box techniques~\cite{Miskin}. In this method, 
the time evolution of the probability distribution of finding the system in configuration $x$ is written as 
   \begin{equation}\label{eq:1}
				\dot{\rho}(x|\lambda_i)=\rho(x|\lambda_i)\left[ f(x)-\langle f(x)\rangle \right],
			\end{equation}
			where $\rho$ denotes the probability of finding a system in some configuration, $\lambda_i's$ are the adjustable parameters and $f(x)$ is a quality function which gives a weight/fitness value to each configuration based on a targeted macroscopic property. 
With straightforward manipulation, the above equation can be recast as equations of motion for $\lambda_i$~\cite{Miskin}:
   \begin{eqnarray}\label{eq:2a}
				\dot{\lambda}_i(t)& =& \langle \partial_{\lambda_i}log\left(\rho\right)\partial_{\lambda_j}log\left(\rho\right)\rangle^{-1} \nonumber \\
				& & \times \langle \left[ f(x)-\langle f(x)\rangle \right]\partial_{\lambda_j}log\left(\rho\right)\rangle,
                \end{eqnarray}
where  $<..>$ denotes an ensemble average at a given set of values of $\lambda_i$.
To integrate Eq.~\ref{eq:2a}, we have used a modified Euler method with a fixed time step of 4.0.\\

We have built our quality function $f(x)$  in the following way:
 \begin{equation}\label{eq:Qlty}
 f(x)=\int dx'\Theta (g(x')\geq g(x))\rho(x'|\lambda)
 \end{equation}
where $g(x)$ denotes the fitness function, 
\begin{equation}
    g(x)=(\bar{q}_6(x)-\bar{q}_6^{target} )^2 + (\bar{w}_6(x)-\bar{w}_6^{target} )^2
\end{equation}
with $\bar{q}_6 (x) = \sum_i^N \bar{q}_6^{(i)} / N$ and $\bar{w}_6 (x)=\sum_i^N \bar{w}_6^{(i)} / N$ the averages of the bond order parameters over all the particles in the system, and $\bar{q}_6^{target}$ and $\bar{w}_6^{target}$ are the corresponding quantities in the target structure (bcc). Here, the integral over $x'$ represents a sum over a series of $n$ different configurations as obtained from a simulation for a fixed set of parameters $(\kappa\sigma, \beta P\sigma^3,\beta\epsilon)$. The quality function $f(x)$ will have higher values for those configurations whose $\bar{q}_6$ and $\bar{w}_6$ values are closer to the target values. More precisely, $f(x)$ equals the probability of having a lower value of $g(x)$ than any other configuration drawn randomly from the equilibrium distribution. To target the bcc structure, we have chosen $\bar{q}_6^{target}=0.395$ and $\bar{w}_6^{target}=0.013161$ (average values of $\bar{q}_6$, $\bar{w}_6$ for the bcc structure obtained from the scatter plot shown in Fig.~\ref{fig:q411}). For a perfect bcc structure, these values are, $\bar{q}_6=0.5107$ and $\bar{w}_6=0.013161$, but here we have targeted the bond order parameter values for a finite-temperature bcc structure.
 
\subsection{Covariance Matrix Adaptation-Evolutionary Strategy} In order to implement CMA-ES,  we draw $n$ samples   from a multivariate Gaussian distribution for each generation whose dimension $D$ corresponds to the number of parameters we wish to tune. Subsequently, we evaluate the fitness function $g(x)$ on the generated samples, and we pick the best $k$ samples. Using the following equations, we estimate the multivariate  Gaussian distribution with mean $\vec \mu $ (a $D$-dimensional vector) and $\mathbf{\Sigma} = \sigma^2\mathbf{C}$  the covariance matrix of the Gaussian distribution for the next generation using:\\
\begin{equation}
\begin{aligned}
{\mu_i}^\prime &= {\mu_i} + \sum_{x}w(x)(\lambda_i (x) - \mu_i) \\
{q_i}^\prime &= (1-c_1)q_i + c_2{\sqrt{\Sigma^{-1}}}_{ij}({\mu_j}^\prime - \mu_j)\\
{p_i}^\prime &= (1-c_3)p_i + c_4({\mu_i}^\prime - \mu_i)\\
{C_{ij}}^\prime &= (1-c_5-c_6)C_{ij} + \\ 
 \quad &+ c_5\sum_{x}w(x)(\frac{\lambda_i (x) - \mu_i}{\sigma}\frac{\lambda_j (x) - \mu_j}{\sigma} - C_{ij} ) + c_6{p_i}^\prime{p_j}^\prime\\
\sigma^\prime &= \sigma \mathrm{\, exp\,}[c_7(\frac{\parallel{q_i}^\prime \parallel}{\langle \parallel N(0,I)\parallel \rangle} - 1)]
\end{aligned}
\end{equation}
where ${\{x\}}$ denotes the $n$ samples consisting of multiple configurations calculated for $n$ different parameter sets $(\kappa\sigma, \beta P\sigma^3,\beta\epsilon)$ (denoted by $\lambda_i(x)$ above) in CMA-ES, $w(x)$ is the normalized distribution of weights based on the fitness of the samples. We choose $w(x) \propto \log(k+1) - \log(i)$ where $i$ is the rank index of sample $x$ ($i = 1$ for the configuration with the smallest $g(x)$ value) for the best $k$ samples, and set $w(x) = 0$ for the rest. $\vec q $ and $\vec p $ are additional $D$-dimensional vectors which determine, respectively, the changes in amplitude and directionality of the covariance matrix, and finally $ \langle \parallel N(0,I)\parallel \rangle $ is the average length of a vector drawn from a multivariate Gaussian distribution centered in the origin and where the covariance matrix is the identity matrix. In the present work we use $n = 10$ and $k = 5$. For the first generation we initialize $\vec q $ and $\vec p $ as null vectors. Moreover, since we do not assume any \emph{a priori} correlation between the different tuning parameters, the initial form of the covariance matrix $\mathbf{\Sigma}$ is diagonal. Finally, all the free parameters of the CMA-ES are selected following the recipe in Ref.~\onlinecite{Hansen2006}. \\

Note that there is a substantial difference in the use of the fitness function between the two methods. In SP-ID we rank different configurations, so $ \bar{q}_6(x) $ and $ \bar{w}_6(x) $ are bond order parameters computed in a single configuration. The SP-ID method is thus based on the statistical fluctuations in the bond order parameters (Fig.~\ref{fig:q411}) in a simulation at a single set of interaction parameters  in order to optimize these values for the desired bcc structure. When using the CMA-ES, we rank samples, so $ \bar{q}_6(x) $ and $ \bar{w}_6(x) $ are computed as ensemble averages of these bond order parameters over multiple configurations for distinct sets of parameters. The CMA-ES method is thus based on a ranking of the  different samples as obtained for  different interaction parameter sets in order to optimize the parameter values. 



\figTwo
\subsection{Simulation details} In order to evaluate the ensemble averages and to generate  distinct configurations for the SP-ID method, we perform constant pressure and constant temperature ($NPT$) Monte Carlo (MC) simulations on systems  consisting of $N=250$ hard-core repulsive Yukawa particles. We initialize the simulations by placing the particles randomly in a cubic simulation box.  We equilibrate the system up to $10^5$ MC cycles. One Monte Carlo cycle corresponds to $N$ particle moves and one volume move. The particle and volume  moves are adjusted in such a way that 45\% of the particle moves and 20\% of the volume moves are accepted. We save $10^3$ uncorrelated samples and evaluate the ensemble averages in Eq.~\ref{eq:2a}. These uncorrelated samples are used to calculate the new parameters. Once the parameters have been changed, we repeat the whole procedure starting the simulation from the last configuration generated with the previous parameters. We repeat this process until the target structure is reached. \\ 

In the CMA-ES algorithm, we evaluate the ensemble averages by performing simulations at different parameter sets at each generation.  At every next generation, we take the last configuration of the fittest sample as  the starting point for all the new samples. \\



 
\figThree
 
\section{Results}\label{sec:3}
\subsection{Tuning parameters with the SP-ID method}
We start from the case in which we tune only one interaction parameter, the inverse Debye screening length $1/\kappa\sigma$ of the interaction potential (Eq.~\ref{eq:pot}) to target the bcc structure. At this stage, pressure $\beta P\sigma^3$ and the contact value $\beta\epsilon$ are kept constant at $\beta P\sigma^3=33$ and $\beta\epsilon=8$. The initial value of $1/\kappa\sigma$ is 0.4, to make sure the system starts from a fluid configuration. Given the form discussed in Sec.~\ref{sec:ID}, the quality function gives higher weights to those configurations whose $\bar{q}_6$ and $\bar{w}_6$ values are closer to the target values. When only one parameter is tuned (1$D$ case), the equation of motion (Eq.~\ref{eq:2a}) becomes,

\begin{equation}\label{eq:e1}
 \frac{d}{dt} (\kappa\sigma) = -Cov[\frac{\partial H}{\partial (\kappa\sigma)},\frac{\partial H}{\partial (\kappa\sigma)}]^{-1}Cov[\frac{\partial H}{\partial (\kappa\sigma)},f],
\end{equation}
where $H=U+PV$ is the Hamiltonian of the system. 
\figFour
By solving Eq.~\ref{eq:e1}, a new value of $1/\kappa\sigma$ is obtained and the algorithm keeps on optimizing this interaction parameter until the goal is reached, i.e.,  $\bar{q}_6=\bar{q}_6^{target}$ and $\bar{w}_6=\bar{w}_6^{target}$. The path of the parameters is shown in Fig.~\ref{denFluid132}(a,b) in the $(1/\kappa\sigma-P)$ and $(1/\kappa\sigma-\eta)$ planes. In both, the optimizer correctly tunes $1/\kappa\sigma$ to reach the bcc structure which can also be verified by examining the evolution of the average $\bar{q}_6$ and $\bar{w}_6$ values as the simulation proceeds. In Fig.~\ref{fig:bop}(a,b), we plot the average $\bar{q}_6$ and $\bar{w}_6$ values as a function of the simulation time. At the very beginning of the simulation, $\bar{q}_6$ and $\bar{w}_6$ values show that the system is in the fluid phase and as the algorithm optimizes the interactions, there is a sharp transition in both of these values which exactly happens at the fluid-bcc phase boundary. Once the system reaches the bcc phase, it remains in the bcc phase. We also find that as the system reaches the phase boundaries, there is a sudden change in the slope of the parameter's trajectory as shown in Fig.~\ref{denFluid132} (c). In other words, the optimizer (Eq.~\ref{eq:2a}) correctly recognizes the phase boundaries present in the phase diagram.
\figFive

We now analyze the case in which we tune two parameters simultaneously (2$D$ case), the inverse Debye screening length $1/\kappa\sigma$ and the pressure $\beta P\sigma^3$ of the interaction potential (Eq.~\ref{eq:pot})  while  $\beta\epsilon=8$ is kept constant. Here, we initialize the system again in the fluid phase at $1/\kappa\sigma=0.4$ and $\beta P\sigma^3=33$. The equations of motion (Eq.~\ref{eq:2a}) for the two parameters  become
\begin{widetext}
\begin{equation}\label{eq:e2}
\frac{d}{dt}\left[ {\begin{array}{c} \beta P \sigma^3 \\ \kappa\sigma\end{array} } \right]=
  -\left[ {\begin{array}{cc}
     Cov[\frac{\partial H}{\partial (\beta P\sigma^3)}, \frac{\partial H}{\partial (\beta P\sigma^3)}] & Cov[\frac{\partial H}{\partial (\beta P\sigma^3)}, \frac{\partial H}{\partial (\kappa\sigma)}]\\
     Cov[\frac{\partial H}{\partial (\beta P\sigma^3)}, \frac{\partial H}{\partial (\kappa\sigma)}] & Cov[\frac{\partial H}{\partial (\kappa\sigma)}, \frac{\partial H}{\partial (\kappa\sigma)}]\\
  
  \end{array} } \right]^{-1} \left[ {\begin{array}{c} 
  
     Cov[\frac{\partial H}{\partial (\beta P\sigma^3)},f] \\
      Cov[\frac{\partial H}{\partial (\kappa\sigma)},f] \\
  
  \end{array} } \right].
\end{equation}
\end{widetext}

In Fig.~\ref{fig:a09}(a,b), we  show the path of the parameters in the $(1/\kappa\sigma-P)$ and $(1/\kappa\sigma-\eta)$  planes. As the simulation time proceeds, SP-ID successfully optimizes both the interaction parameters in such a way that the final structure formed is the bcc crystal. The form of the quality function is the same as we have used for the one parameter case. Variations of $\bar{q}_6$ and $\bar{w}_6$ values also verify the formation of the bcc structure from the fluid phase (Fig.~\ref{fig:bop}(c,d)). We find from Fig.~\ref{fig:a09}(c) that the optimizer recognizes the phase boundaries very well as also found for the one parameter case. 

\figSix

Finally, we investigate the case in which we tune three parameters simultaneously (3$D$ case), the inverse Debye screening length $1/\kappa\sigma$, the pressure $\beta P\sigma^3$ and the  contact value $\beta\epsilon$ of the interaction potential (Eq.~\ref{eq:pot}). We perform two independent simulations by initializing the system at two different state points in the fluid phase at (i)
$1/\kappa\sigma=0.4$, $\beta P\sigma^3=33$ and $\beta\epsilon=8$, (ii) $1/\kappa\sigma=0.4$, $\beta P\sigma^3=25$ and $\beta\epsilon=6$ and optimize all  three parameters, $\kappa\sigma$, $\beta P\sigma^3$ and $\beta\epsilon$ for the bcc phase. The equations of motion (Eq.~\ref{eq:2a}) when three parameters are tuned become
\begin{widetext}
\begin{equation}
\frac{d}{dt}\left[ {\begin{array}{c} \beta P\sigma^3 \\ \kappa\sigma \\ \beta\epsilon\end{array} } \right]=
  -\left[ {\begin{array}{ccc}
     Cov[\frac{\partial H}{\partial (\beta P\sigma^3)}, \frac{\partial H}{\partial (\beta P\sigma^3)}] & Cov[\frac{\partial H}{\partial (\beta P\sigma^3)}, \frac{\partial H}{\partial (\kappa\sigma)}] & Cov[\frac{\partial H}{\partial (\beta P\sigma^3)}, \frac{\partial H}{\partial (\beta \epsilon)}]\\
      Cov[\frac{\partial H}{\partial (\kappa\sigma)}, \frac{\partial H}{\partial (\beta P\sigma^3)}] & Cov[\frac{\partial H}{\partial (\kappa\sigma)}, \frac{\partial H}{\partial (\kappa\sigma)}] & Cov[\frac{\partial H}{\partial (\kappa\sigma)}, \frac{\partial H}{\partial (\beta \epsilon)}]\\
      Cov[\frac{\partial H}{\partial (\beta\epsilon)}, \frac{\partial H}{\partial (\beta P\sigma^3)}] & Cov[\frac{\partial H}{\partial (\beta\epsilon)}, \frac{\partial H}{\partial (\kappa\sigma)}] & Cov[\frac{\partial H}{\partial (\beta\epsilon)}, \frac{\partial H}{\partial (\beta \epsilon)}]\\

  \end{array} } \right]^{-1} \left[ {\begin{array}{c} 
  
     Cov[\frac{\partial H}{\partial (\beta P\sigma^3)},f] \\
      Cov[\frac{\partial H}{\partial (\kappa\sigma)},f] \\
      Cov[\frac{\partial H}{\partial (\beta\epsilon)},f] \\
  
  \end{array} } \right].
\end{equation}
\end{widetext}


In Fig.~\ref{fig:13}, we plot the path of the tuned parameter trajectories as a function of simulation time when the system is initialized in the  fluid phase and the desired goal is to reach the targeted bcc structure. Initially, all three parameter values decrease while the system is in the fluid phase and once it crosses the phase boundary between the fluid and bcc phase, they start to  saturate.
As the simulation time proceeds, the optimizer successfully optimizes  the  parameters in such a way that the final structure becomes the bcc phase. Here we also use the same form of the quality function as we have used earlier  for the one and two parameter cases.\\

In Table~\ref{table7}, we show the state point values from which the optimization starts, which correspond to a fluid phase, as well as the final state point obtained from the optimization algorithm, which corresponds to the bcc phase. Note that the parameter values of the pressure and screening length are different  as compared to the one- and two-parameter cases, because the contact value $\beta \epsilon$ is also being optimized.  

\tableOne

To confirm that the final structure is a bcc phase, we also plot the evolution of $\bar{q}_6$ and $\bar{w}_6$  as a function of the simulation time in Fig.~\ref{fig:bop}(e,f), which indeed confirms that the simulation reaches the optimal bond order parameter values  for the bcc phase. 

\subsection{Tuning parameters with the CMA-ES method}

We now employ the CMA-ES algorithm to analyze all the cases already studied using the SP-ID method, i.e. the tuning of one, two and three parameters, highlighting the differences between the two inverse design optimizers. We use the parameters corresponding to the initial state point employed in the SP-ID algorithm  as the initial mean vectors of the multivariate Gaussian distribution in the CMA-ES algorithm. In addition, we start the CMA-ES algorithm with a diagonal covariance matrix with a standard deviation of 10\% around its mean value for  each parameter. Finally, the initial values of each component of the vectors $\vec q$ and $\vec p$ are set to zero. \\

\figSeven

In Fig. ~\ref{fig:cma1} we show the results for the one parameter case. The points displayed in Fig~\ref{fig:cma1}(a) represent the mean value of the Gaussian distribution in each generation. At the beginning, the algorithm tries to decrease the fitness function value by decreasing the inverse Debye screening length $1/k\sigma$. Since all the steps point in the same direction, the variance of the Gaussian distribution increases  (Fig~\ref{fig:cma1}(b)) and the mean value overtakes the bcc region, ending up in the fcc phase. After this, the algorithm recognizes the right direction, and  the subsequent update is in the opposite direction, i.e. the inverse Debye screening length increases. At the same time, the covariance starts to shrink and the mean value of the Gaussian distribution is found in the bcc region for the first time at the $6^{th}$ generation. From this generation onwards the updates are in random directions inside the bcc region, leading to a further exponential decrease of the covariance of the Gaussian distribution. At the $12^{th}$ generation, all the 10 simulations have a $k\sigma$ value for which the structure corresponds to the stable bcc phase. \\

\figEight
\figNine
\figTen
The cases in which we tune two or three parameters do not present significantly different behaviour of the CMA-ES algorithm with respect to the one parameter case. In the two parameter case, $\vec \mu$ enters   the bcc region for the first time  at the $10^{th}$ generation, while at the $18^{th}$ generation all the samples have entered the bcc region, showing how CMA-ES is adversely affected by the dimensionality of the parameter space. The results for the 2$D$ case are shown in Fig~\ref{fig:cma2}. We note that the risk of overshooting is lower when more parameters are varied at the same time, since they all contribute to the increase or decrease of the quality function, and the updates of the mean values of the multivariate Gaussian distribution may therefore not always be in the same direction. This prevents  the covariance matrix from growing too fast as already shown. Finally, the results for both the investigated 3$D$ cases are shown in Fig~\ref{fig:cma3}. We stress that, when $\beta \epsilon$ varies, the only way to know if we are inside the bcc region is to plot the bond order parameters and check their values, as we show in Fig~\ref{fig:cmabop}. \\

These analyses provide benchmarks for how fast the CMS-ES algorithm converges in finding the target structure, as compared to the SP-ID algorithm. Since each generation of the CMA-ES requires  $n$ times (10 in this work) the computational effort of SP-ID, a comparison of run lengths in the two methods reveals that their efficiencies are very similar. However, in SP-ID, one has information on when the phase boundary is crossed.  In CMA-ES the size of the steps from one generation to the other varies, depending on the current form of the covariance matrix. This makes the algorithm explore the landscape in an optimal way, sacrificing information about the phase boundaries. Thus, we conclude that SP-ID and CMA-ES are comparable in their performance, but SP-ID has an edge in retrieving information on the phase boundaries.   

\section{Conclusions}\label{sec:4}
We studied the inverse problem of tuning interaction parameters between charged colloids interacting via a hard-core repulsive Yukawa potential, so that they self-assemble into a targeted crystal structure. We targeted the bcc structure which occupies a narrow region in the phase diagram of the above system and is therefore challenging to find. We showed a comparison between two different optimization algorithms  in order to achieve our goal: Statistical Physics-inspired Inverse Design (SP-ID) and Covariance Matrix Adaption - Evolutionary Strategy (CMA-ES). The first makes use of the statistical fluctuations in the bond order parameters to iteratively change the interaction parameters of the system. In addition to effectively tuning the interaction parameters  for obtaining the target structure, the SP-ID method  correctly identifies the fluid-solid phase boundaries present in the phase diagram. The CMA-ES algorithm generates samples from a multivariate Gaussian distribution at each generation and evaluates the fitness of these samples in order to evolve the interaction parameters of the distribution. The number of generations needed to reach the goal is on average lower in the case of the CMA-ES, and the steps in parameter space are usually larger. This advantage is offset by the need to simulate multiple samples, and we find that the computational effort required in the two methods is comparable. On the other hand, because of the larger step sizes of the parameters, probing phase equilibrium with CMA-ES can be less straightforward than with SP-ID. Most importantly,  we showed that both of these inverse methods lead to the targeted bcc structure by tuning the interactions between the particles. Thus, our results demonstrate both methods to be effective search algorithms that may be employed in other design tasks. Although the quality function used here is strictly structural, one may in principle also include quantifiers of dynamics, which may be useful in optimizing the kinetic self-assembly pathways, for example, to account for and exclude glassy dynamics, as well as other kinetic factors of self assembly. 

\acknowledgments{We gratefully acknowledge the India-Netherlands (DST-NWO) joint project grant for supporting the collaboration between the authors. }

\bibliography{main} 
 \bibliographystyle{prsty}
\end{document}